\documentclass[hyper]{JHEP} 

\usepackage{epsfig}

\newcommand\fverb{\setbox\pippobox=\hbox\bgroup\verb}
\newcommand\fverbdo{\egroup\medskip\noindent%
			\fbox{\unhbox\pippobox}\ }
\newcommand\fverbit{\egroup\item[\fbox{\unhbox\pippobox}]}
\newbox\pippobox
\title{$N$ D0-branes and $\overline{D}$0-branes in
the background independent string field theory.}

\author{by J. Kluso\v{n}\\
	 Department of Theoretical Physics and Astrophysics\\
                   Faculty of Science, Masaryk University\\
Kotl\'{a}\v{r}sk\'{a} 2, 611 37, Brno\\
Czech Republic\\
	E-mail: \email{klu@physics.muni.cz}}

\preprint{\hepth{0103079}}	

\abstract{In this paper we extend our previous
work \hepth{0102063} to the  case of $D\overline{D}$0-brane
system.}


\def\tr{\mathrm{Tr}}

\def\ket #1{\left|#1\right>}
\def\dd{$D\overline{D}$}
\begin{document}

\section{Introduction}\label{first}

Around the year '92-'93  the formulation
of the string field theory known as a background
independent string field theory (BSFT) 
\cite{WittenBT,WittenBT1,WittenBT2,ShatasviliBT,
ShatasviliBT1} was proposed. 
In the recent papers 
\cite{ShatasviliBT2, MooreBT, 
SenBT, Cornalba, Okuyama, MooreBT2,
Aleksejev,DasguptaBT1,Moriyama,
ShatasviliBT3,Yasnov,Kleban,KrausBST,TakayBT,Takayanagi}
 (For the world-sheet
approach to this problem, see \cite{TseytlinSM,HarveyWS1,
HarveyWS2})
 the BSFT was used in the analysis
of the tachyon condensation on the unstable D-brane
systems
in string theories (For review and extensive list of 
references, see \cite{SenR}). It was shown that BSFT is
very effective tool for the study of this problem
since in the process of the tachyon condensation 
only tachyon field acquires  nonzero
vacuum  expectation value. 
 In particular,
it was shown that:
\begin{itemize}
\item The condensation to the closed string
vacuum and to lower dimensional branes involves excitations
of only one mode of string field-tachyon.
\item The exact tachyon potential can be computed in BSFT and
its qualitative features agree with the Sen conjecture
\cite{SenC}.
\item The exact tachyon profiles corresponding to the tachyon
condensation to the lower dimensional D-brane give rise to
descent relations between the tensions of various branes 
\cite{Moriyama} which again agree with those expected
\cite{SenC}. 
\end{itemize}
The problem of tachyon condensation was also studied from
the point of view of Witten's open string field theory 
\cite{WittenSFT} (For recent discussion see \cite{SenSF1,SenSF2}
and for the review and extensive list of references, see \cite{Ohmori}).
 In contrast
with BSFT, the tachyon condensation in general involves 
 giving the expectation values of infinite number of component fields.
As a consequence, only approximate results are available. 

In \cite{KlusonBS} the problem of the tachyon condensation
was  studied from  different point of view
\footnote{The same problem has been already studied in
terms of the low energy effective action in \cite{Kluson1} and
in the framework of the effective action obtained from
BSFT theory in \cite{Terashima}.}. Namely, we started with the
system of $N$ D-instantons in bosonic theory or with the
system of $N$ non-BPS D-instantons in Type IIA theory and
studied the emergence of higher dimensional D-branes
in this system. We have shown that all even dimensional D-brane
arise naturally through the standard matrix theory construction
\cite{TaylorM1,TaylorM2,
Schwarz,TaylorM3} . However to describe odd dimensional D-branes
the tachyon field should be turned on and then in the process
of the tachyon condensation \cite{MooreBT,MooreBT2} all odd dimensional 
D-branes emerged. 

In this paper we extend this approach to the case
of the system of $N$ D0+$\overline{D}0$-branes (\dd0-system.) 
The general \dd \ system has been recently studied in  \cite{KrausBST,TakayBT}
and we use results obtained there in our calculations.
 In fact, \dd \ system
seems to contain much  more information about the
tachyon condensation than non-BPS D-branes
since  all non-BPS D-branes can arise as a particular projection
from \dd \ system \cite{SenR}. For that reason it seems to be natural
to ask the question whether it is possible to describe all D-branes from 
$N$ \dd0 branes in type IIA theory. We will show in this paper that
 this is really possible.

The organisation of the paper is as follows. In the next section
we present the partition function on the disk
 for $N$ \dd 9-branes which, thanks to the recent conjecture
\cite{MooreBT2}, should be identified with the space-time action
in BSFT. Our form of the boundary interaction
can be considered as a slight modification of the action given
in \cite{KrausBST}. In order to show that this boundary term
gives correct results we will argue that the action for \dd \ system
 reduces to
the action for D9-branes without the tachyon field and to
the action for non-BPS D9-brane after orbifolding projection
\cite{SenR}.

In (\ref{third}) section we will 
 review the approach presented in
\cite{KlusonBS} on the example
of $N$ non-BPS D0-branes. 

In (\ref{fourth}) section we will turn to the main theme
of this paper that is the emergence of all D-branes
from the system of $N$ \dd 0-branes in type IIA theory.
We will show that all Dp-branes  naturally
emerge either directly in the standard matrix theory
construction or through the tachyon condensation
worked out in \cite{KrausBST,TakayBT}. 
We will also discuss other possibilities of the emergence
of D-branes that are based on nontrivial gauge fields
on the world-volume of \dd  \ system. Unfortunately,
this possibility does not seem to lead to the free boundary
theory so that it is difficult  to study it in the BSFT theory.
 
In conclusion (\ref{fifth}) we will discuss some
open problems and possible extension of this work.

\section{\dd -system}\label{second}
In this section we will consider
 \dd  \ system
 from the point of view of BSFT following recent
 papers \cite{KrausBST,TakayBT}. 
The general construction of the BSFT for $N$ D-branes
and antibranes was given in \cite{KrausBST} which we 
will closely follow. The starting point  of the BSFT
for \dd \ system is  the partition function 
on the disk
\begin{equation}\label{S}
S(A,A',T)=Z(A,A',T)=
<\tr \hat{P}\exp\left(\int_0^{2\pi}
d\hat{\tau}M(A,A',T) \right)> \ ,
\end{equation}
where the symbol $\hat{P}$ is defined as
\begin{equation}
\hat{P}e^{\int d\hat{\tau} M(\hat{\tau})}=
\sum_{N=0}^{\infty}
\int d\hat{\tau}_1\dots d\hat{\tau}_N
\Theta(\hat{\tau}_{12})\Theta
(\hat{\tau}_{23})\dots \Theta(\hat{\tau}_{N-1,N})
M(\hat{\tau}_1)M(\hat{\tau_2})\dots
M(\hat{\tau}_N) \  ,
\end{equation}
and where $d\hat{\tau}=
d\tau d\theta, \ \hat{\tau}_{12}=\tau_1-\tau_2-\theta_1\theta_2$
and $\Theta$ is a step function whose expansion is
equal to $\Theta(\hat{\tau}_1-\hat{\tau}_2)=
\theta(\tau_1-\tau_2)-\delta(\tau_1-\tau_2)
\theta_1\theta_2$. 
The matrix $M$ describes general
boundary interaction. In situation, when
we confine to the gauge fields and  tachyon,
it  has a form \cite{KrausBST}
\begin{equation}\label{M1}
M(A,A',T)=
\left(\begin{array}{cc}
-iA_{\mu}({\bf X})D{\bf X}^{\mu} &
\overline{T}({\bf X}) \\
T({\bf X}) & -iA'_{\mu}({\bf X})D{\bf X}^{\mu}
\\ \end{array}\right) ,  \ {\bf X}^{\mu}=
X^{\mu}+\sqrt{\alpha'}\theta \psi^{\mu} \ ,
\end{equation}
where $A_{\mu}$   is a gauge field in the adjoint 
representation of the gauge group $U(N)$ living on the world-volume
of 
$N$ D9-branes, $A'_{\mu}$ is a  gauge
field in the adjoint representation of the gauge
group $U(N)$ 
living on the world-volume of $N$ 
$\overline{D}$9-branes and $T$ is
a tachyon field in  $(N,\overline{N})$ representation
of the gauge group 
$U(N)\times U(N)$
\footnote{For simplicity we will consider
the configurations with the equal number  branes
and antibranes. The generalisation to the more
general case is trivial.}. The previous form
of the boundary interaction corresponds to
the system of $N$ D9-branes and antibranes, generalisation
to the lower dimensional case is straightforward.
 In (\ref{M1}) we presume that
off-diagonal terms have odd grading. In other words,
the form of (\ref{M1}) has a property similar to the
superconnection \cite{Quillen} (For detailed discussion of the
relation between superconnection and BSFT,
see  \cite{KrausBST,TakayBT,Oz}). 
In order to show that (\ref{M1}) defines correctly
BSFT action for \dd9-system we express the 
partition function in the more familiar form
using elegant formalism reviewed in \cite{TseytlinSM}
which allows us to rewrite the path-ordered trace in
terms of the integral over fermionic fields 
\begin{equation}
S=<\exp \left(\int d\hat{\tau} \left[
\hat{\overline{\eta}}_a D \hat{\eta}^a+
\hat{\overline{\eta}}_aM^a_b\hat{\eta}^b
\right]\right)>=<e^{-I}> \ , 
\end{equation}  
where we have introduced fermionic superfields
$\overline{\hat{\eta}}_a
=\overline{\eta}_a+\theta\overline{\chi}_a,
\hat{\eta}^a=\eta^a+\theta \chi^a, \ a=1,\dots, 2N$
living on the boundary of the disk. Firstly we perform the
integration over
$\theta$. In order to do  that we  expand $M$  
 as $M=M_0+\theta M_1$
with
\begin{equation}
M_0=\left(\begin{array}{cc}
-i\sqrt{\alpha'}A_{\mu}\psi^{\mu} 
& \overline{T}(X) \\ 
T(X) & 
-i\sqrt{\alpha'}A'_{\mu}\psi^{\mu}
\\ \end{array}\right) \ , 
\end{equation}
and 
\begin{equation}
M_1=\left(\begin{array}{cc}
-i(A_{\mu}(X)\partial_{\tau}X^{\mu}+\frac{\alpha'}{2}
(\partial_{\mu}A_{\nu}-\partial_{\nu}A_{\mu})
\psi^{\mu}\psi^{\nu}) & 
-\partial_{\mu}\overline{T}(X)\sqrt{\alpha'}\psi^{\mu} \\ 
-\partial_{\mu}T(X)\sqrt{\alpha'}\psi^{\mu}  & 
-i(A'_{\mu}(X)\partial_{\tau}X^{\mu}+\frac{\alpha'}{2}
(\partial_{\mu}A'_{\nu}-
\partial_{\nu}A'_{\mu})\psi^{\mu}\psi^{\nu}) \\ \end{array}\right) \ .
\end{equation}
Then  
\begin{equation}
-\int d\tau d\theta
\overline{\hat{\eta}}_aM^a_b\hat{\eta}^b=
\int d\tau (-\overline{\eta}_aM^a_{0b}\chi^b-
\overline{\chi}_aM^a_{0b}\eta^b+
\overline{\eta}_aM^a_{1b}\eta^b) \ .
\end{equation}
In the previous expression we have used the crucial
fact that $M_0$ anticommutes with $\theta$ which
is a consequence of the presence of the 
fermionic field $\psi$ on the main diagonal of $M_0$ 
and the presumption that
off-diagonal elements have odd grading so that 
anticommute with fermionic fields.
 We must also stress
that in $M_1$ the off-diagonal terms have a even grading
thanks to odd grading of $T$ and odd grading of $\psi$.
We then obtain  
\begin{equation}\label{I11}
I=\int d\tau\left(
\overline{\eta}_a
\partial_{\tau}\eta^a-\overline{\chi}_a\chi^a 
-\overline{\eta}_aM^a_{0b}\chi^b-
\overline{\chi}_aM^a_{0b}\eta^b+
\overline{\eta}_aM^a_{1b}\eta^b\right)  \ .
\end{equation}
We can integrate out the auxiliary fields $\overline{\chi}_a,
\chi^a$ with the result
\begin{equation}
\overline{\chi}_b=-\overline{\eta}_aM^a_{0b} \ ,
\chi^a=-M^a_{0b}\eta^b \ ,
\end{equation}
and consequently (\ref{I11}) is equal to
\begin{equation}\label{I12}
I=\int d\tau \left(
\overline{\eta}_a
\partial_{\tau}\eta^a+\overline{\eta}_a(
M^a_{0c}M^c_{0b}+M^a_{1b})\eta^b\right) \ ,
\end{equation}
with 
\begin{equation}
M^a_{0c}M^c_{0b}=(M_0^2)^a_b=
\left(\begin{array}{cc}
-\frac{\alpha'}{2}[A_{\mu},A_{\nu}]
\psi^{\mu}\psi^{\nu} +\overline{T}T &
-i\sqrt{\alpha'}(\overline{T}A'_{\mu}-A_{\mu}
\overline{T})\psi^{\mu} \\
-i\sqrt{\alpha'}(TA_{\mu}
-A'_{\mu}T)\psi^{\mu} & -\frac{\alpha'}{2}
[A_{\mu}',A'_{\nu}]\psi^{\mu}\psi^{\nu}
+T\overline{T} \\
\end{array}\right) \ .
\end{equation}
We must stress that
 odd grading  of  off-diagonal elements
($T$) is crucial for the emergence of the
correct form of the covariant derivative.
Then   (\ref{I12}) is equal to
\begin{eqnarray}\label{I13}
I=\int d\tau \left(\overline{\eta}_a
\partial_{\tau}\eta^a+\overline{\eta}_a
\mathcal{M}^a_b\eta^b\right) \ , 
\mathcal{M}^a_b=M_{1b}^a+(M_0^2)^a_b \ , \nonumber \\
\mathcal{M}=
\left(\begin{array}{cc}
-i(A_{\mu}\partial_{\tau}X^{\mu}+
\frac{\alpha'}{2}F_{\mu\nu}\psi^{\mu}\psi^{\nu}) 
+\overline{T}T&
-\sqrt{\alpha'}\overline{D_{\mu}T}\psi^{\mu}  \\
-\sqrt{\alpha'}D_{\mu}T\psi^{\mu} &
-i(A'_{\mu}\partial_{\tau}X^{\mu}+
\frac{\alpha'}{2}F'_{\mu\nu}\psi^{\mu}\psi^{\nu})
+T\overline{T} \\ \end{array}\right) \ , \nonumber \\
\end{eqnarray}
with
\begin{eqnarray}
F_{\mu\nu}=\partial_{\mu}A_{\nu}-
\partial_{\nu}A_{\mu}-i[A_{\mu},A_{\nu}], \
F_{\mu\nu}'=\partial_{\mu}A'_{\nu}-
\partial_{\nu}A'_{\mu}-i[A'_{\mu},
A'_{\nu}] \ , \nonumber \\
D_{\mu}T=\partial_{\mu}T+i(TA_{\mu}-
A'_{\mu}T), \ 
\overline{D_{\mu}T}=\partial_{\mu}\overline{T}
+i(\overline{T}A_{\mu}'-A_{\mu}\overline{T}) \ .
\nonumber \\
\end{eqnarray}
As in \cite{TakayBT} we perform some simple tests
that will show that (\ref{I13}) could really be the correct
form of the action for \dd \ system.
 Firstly, for $T=0$ we expect that the 
partition function will be equal to the
sum of two independent partition functions
for $N$ D9-branes and $N$ $\overline{D}9$-branes. In this
case  $\mathcal{M}$ is block diagonal and when we
express the integral over $\eta,\overline{\eta}$ in terms of
the path-ordered trace then the
partition function is a sum of
two independent terms corresponding to $N$ D9-branes and $N$
$\overline{D}$9-branes respectively
\begin{equation}
S_{D+\overline{D},T=0}
=\sum_{D,\overline{D}}
<\tr P\exp\left(i\int d\tau\left[ 
A_{\mu}(X)\partial_{\tau}X^{\mu}+
\frac{\alpha'}{2}F_{\mu\nu}\psi^{\mu}\psi^{\nu}\right]
\right)> \ . 
\end{equation}
The more nontrivial task is to show that (\ref{I13}) reduces
to the partition function for a non-BPS D9-brane through
the orbifolding projection
$(-1)^{F_L}$ \cite{SenR} that acts on the
tachyon and gauge fields as
\begin{eqnarray}
T=\overline{T}, A_{\mu}=A_{\mu}'
\Rightarrow D_{\mu}T=\partial_{\mu}T
-i[A_{\mu},T], \ \overline{D_{\mu}T}=
\partial_{\mu}T-i[A_{\mu},T] \ , \nonumber \\
\mathcal{M}=\left(\begin{array}{cc}
-i(A_{\mu}\partial_{\tau}X^{\mu}+
\frac{\alpha'}{2}F_{\mu\nu}\psi^{\mu}\psi^{\nu}) 
+TT&
-\sqrt{\alpha'}D_{\mu}T\psi^{\mu}  \\
-\sqrt{\alpha'}D_{\mu}T\psi^{\mu} &
-i(A_{\mu}\partial_{\tau}X^{\mu}+
\frac{\alpha'}{2}F_{\mu\nu}\psi^{\mu}\psi^{\nu})
+TT \\ \end{array}\right) \ . \nonumber \\
\end{eqnarray}
We again rewrite the path integral
over $\eta,\overline{\eta}$ as a path-ordered trace
and then  express $\mathcal{M}$ as
\begin{equation}
\mathcal{M}=1_{2\times 2}\otimes
\mathcal{M}_1 +
\sigma_1\otimes\mathcal{M}_2,  \
\sigma_1=\left(\begin{array}{cc}
0 & 1 \\
1 & 0 \\ \end{array}\right) \ .
\end{equation}
It is clear that  the
  trace over
$2N\times 2N$ matrices is the same as the
 trace over  $N\times N$ matrices  
$\mathcal{M}_{0,1}$ and the trace over 
$2\times 2$ matrices $1, \ \sigma_1$
\begin{equation}
\tr \rightarrow \tr_N\tr_{2\times 2} \ .
\end{equation}
When we convert the trace over $N\times N$ matrices
into the path integral over fermions it is clear
that now we only need $N$ fermionic fields $\eta^a,
\overline{\eta}_a$. 
To proceed further we must  express the 
$2\times 2$ matrices as  fermionic fields. 
 We will mainly follow \cite{Polchinski}.
Consider the basic fermionic quantum system, two states
$\ket{-},\ket{+}$ with raising operator $\xi$ and lowering
operator $\overline{\xi}$
\begin{equation}
\xi \ket{-}=0, \ \overline{\xi}\ket{-}=\ket{+}, \
\xi\ket{+}=\ket{-}, \ \overline{\xi}\ket{+}=0, \  
\{\xi,\overline{\xi}\}=1, \ \xi^2=\overline{\xi}^2=0 \ . 
\end{equation}
Then immediately follows that 
$\xi,\overline{\xi}$ have a matrix representation
in the basis $\ket{-},\ket{+}$
\footnote{$\sigma_1,\sigma_2$ are ordinary Pauli matrices
$\sigma_1=\left(\begin{array}{cc}
0 & 1 \\
1 & 0 \\ \end{array}\right) , \
\sigma_2=\left(\begin{array}{cc}
0 & -i \\
i & 0 \\ \end{array}\right) $.}
\begin{equation}
\xi=\left(\begin{array}{cc}
0 & 1 \\
0 & 0 \\ \end{array}\right)=\frac{1}{2}
(\sigma_1+i\sigma_2) \ , 
\overline{\xi}=\left(\begin{array}{cc}
0 & 0 \\
1 & 0 \\ \end{array}\right)=\frac{1}{2}(
\sigma_1-i\sigma_2) \ .
\end{equation}
We see that we can replace
 $\sigma_1$ in the partition sum
 with $\sigma_1=\xi+\overline{\xi}$
and  the trace with the path-integral 
over fermionic fields $\xi,\overline{\xi}$.
More precisely, when we insert
$\int d\tau \overline{\xi}\partial_{\tau}{\xi}$ into
the action $I$  
 we obtain
\begin{eqnarray}
I=\int d\tau 
\left(\overline{\eta}_a\partial_{\tau}\eta^a+
\overline{\xi}\partial_{\tau}\xi-i\overline{\eta}_a
\left(A_{\mu b}^a\partial_{\tau}X^{\mu}+
\frac{\alpha'}{2}F_{\mu\nu b}^a\psi^{\mu}\psi^{\nu}
\right)\eta^b+\right. \nonumber \\
+\left. \overline{\eta}_a
T^a_cT^c_b\eta^b-(\xi+\overline{\xi})
\overline{\eta}_a \sqrt{\alpha'}D_{\mu}T^a_b
\psi^{\mu} \eta^b\right) \ . \nonumber \\
\end{eqnarray}
It is important to stress that thanks to the replacement 
$\sigma_1$ with the fermionic operator $\xi+
\overline{\xi}$
 the tachyon field $T(X)$ should be regarded as
a  field with  even grading.
The integration over $\xi,\overline{\xi}$  gives
\begin{equation}
\xi=\sqrt{\alpha'}\frac{1}
{\partial_{\tau}}\left(\overline{\eta}_aD_{\mu}T^a_b
\psi^{\mu}\eta^b\right) \ ,
\overline{\xi}=\sqrt{\alpha'}\frac{1}
{\partial_{\tau}}\left(\overline{\eta}_aD_{\mu}T^a_b
\psi^{\mu}\eta^b\right) \ ,
\end{equation}
so that
\begin{equation}
\overline{\xi}\partial_{\tau}\xi-
(\xi+\overline{\xi})\overline{\eta}D_{\mu}T\psi^{\mu}\eta
=-\alpha'\frac{1}{\partial_{\tau}}
(\overline{\eta}D_{\mu}T\psi^{\mu}\eta)
(\overline{\eta}D_{\nu}T\psi^{\nu}\eta)
=\alpha'(\overline{\eta}
D_{\mu}T\psi^{\mu}\eta)\frac{1}{\partial_{\tau}}(\overline{\eta}
D_{\nu}T\psi^{\nu}\eta) \ , 
\end{equation}
where in the last step we have used the anticommuting
property of fermionic fields. 
Finally,  $S$ is equal to
\begin{eqnarray}\label{nBPS}
S=<e^{-I}>=<\exp\left(-\int d\tau 
\left\{\overline{\eta}_a\partial_{\tau}\eta^a
-i\overline{\eta}_a
\left(A_{\mu b}^a\partial_{\tau}X^{\mu}+
\frac{\alpha'}{2}F_{\mu\nu b}^a\psi^{\mu}\psi^{\nu}
\right)\eta^b+\right.\right. \nonumber \\
+\left.\left. \overline{\eta}_a
T^a_cT^c_b\eta^b+\alpha'\left(\overline{\eta}_a
D_{\mu}T^a_b\psi^{\mu}\eta^b\right)
\frac{1}{\partial_{\tau}}\left(
\overline{\eta}_aD_{\nu}T^a_b\psi^{\nu}\eta^b
\right)\right\}\right)> \ , \nonumber \\
\end{eqnarray}
which is in agreement  with 
the partition function for $N$ non-BPS 
D9-branes given in \cite{MooreBT2,TseytlinSM}.
In the next section we use this 
partition function for $N$ non-BPS D0-branes to
review the results presented in \cite{KlusonBS}.
\section{D-branes from $N$ non-BPS
D0-branes}\label{third}
In this section we will construct 
all D-branes from $N$ non-BPS D0-branes
in type IIB theory in the same way
as in \cite{KlusonBS}. 
We start with the action (\ref{nBPS}).
Using transformation properties of  
the gauge fields and covariant derivatives under T-duality
\begin{eqnarray}
A_I(X^{\mu})\rightarrow \Phi_I(X^0) , \ I,J=1,\dots, 9, \
\partial_{\tau}X^I\rightarrow \partial_n X^I \ ,
T(X^{\mu})\rightarrow T(X^0) \ ,  \nonumber \\
F_{IJ}\rightarrow -i[\Phi_I,\Phi_J], \
F_{0I}\rightarrow D_0\Phi_I, 
D_{I}T=-i[\Phi_I,T] \ , \ \nonumber \\
\end{eqnarray}
we obtain the partition function for $N$ non-BPS
D0-branes.
It is important to stress that all background fields
$A_0,\Phi_I,T$ are functions of $X^0(\tau)$ only since
this is the  string coordinate with the Neumann
boundary condition. In the following we will work
with the time independent background field so
that they are not functions of $X^0$ and we
also use gauge invariance to eliminate
 the gauge field $A_0$.
 Then  the BSFT action
for $N$ non-BPS D0-branes is
\begin{eqnarray}\label{nBPS0}
S(\Phi,T)=<e^{-I}>=
<\exp\left(-\int d\tau 
\left(\overline{\eta}_a\partial_{\tau}\eta^a
-i\overline{\eta}_a
\left(\Phi_{I b}^a\partial_{n}X^{I}-
i\frac{\alpha'}{2}[\Phi_I,\Phi_J]_b^a\psi^{I}
\psi^J
\right)\eta^b+\right.\right. \nonumber \\
+\left.\left. \overline{\eta}_a
T^a_cT^c_b\eta^b-\left(\overline{\eta}_a
[\Phi_I,T]^a_b\psi^{I}\eta^b\right)
\frac{1}{\partial_{\tau}}\left(
\overline{\eta}_a[\Phi_J,T]^a_b\psi^{J}\eta^b
\right)\right)\right)> \ . \nonumber \\
\end{eqnarray}
We will show that the tachyon condensation in
the action given (\ref{nBPS0}) can describe all
D-branes in type IIB theory, following 
\cite{KlusonBS}.  
Let us consider the background configuration
in the form
\begin{equation}\label{backk}
T=0, \ [\Phi_a,\Phi_b]=i\theta_{ab}, \ a=1,\dots, 2p, \ 
\Phi_{\alpha}=0, \ \alpha=2p+1,\dots, 9 \ .
\end{equation}
Then the fermionic term  in 
(\ref{nBPS0}) is equal to
\begin{equation}
-i\frac{\alpha'}{2}[\Phi_a,\Phi_b]\psi^a(\tau)
\psi^b(\tau)=\frac{\alpha'}{2}\theta_{ab}\psi^a(\tau)
\psi^b(\tau)1_{N\times N} \ .
\end{equation}
We see that this term is proportional to the unit matrix so that
the following expression can be taken out the trace
\begin{equation}
\exp\left( i\int d\tau \frac{\alpha'}{2}\theta_{ab}\psi^a 
(\tau)\psi^b(\tau)\right) \ .
\end{equation}
The bosonic part can be worked out  exactly as
in \cite{KlusonBS} and we briefly review this calculation.
From (\ref{backk}) we see that $\Phi_a$ are equivalent to
the quantum mechanics operators with nontrivial commuting
relations. Then we can use the well known relation between
the trace over Hilbert space and the path integral
\begin{equation}
\tr P\exp\left (-i\int d\tau H(\tau)\right)=
\exp\left( i\int d\tau\left[p(\tau) \dot{q}(\tau) -H(p(\tau),q
(\tau)) \right]\right) \ , 
\end{equation}
with $\dot{q}=\partial_{\tau} q$ and where $p$ is a momentum
conjugate to $q$. In our case, the Hamiltonian is
\begin{equation}
H(\tau)=-\Phi_a\partial_nX^a(\tau) \ .
\end{equation}
with the  operators $\Phi_a$. Then
we can rewrite the trace in (\ref{nBPS0}) as a path integral
(For the time being we omit the fermionic term which
is proportional to the unit matrix)
\begin{equation}
\int \prod_{a=1}^{2p}
[\phi_a]\exp\left(i\int d\tau
(\frac{1}{2}\phi_a(\tau)\theta^{ab}\dot{\phi}_b(\tau)
+\phi_a(\tau)\partial_nX^a (\tau))\right) \  ,
\end{equation}
with $\theta_{ac}\theta^{cb}=\delta_a^b$.
We can easily perform the integration over $\phi$
\begin{eqnarray}\label{Jef}
\int [d\phi_a]\exp\left(-\int d\tau d\tau' \left[
\frac{1}{2}\phi_a(\tau')\triangle(\tau',\tau)^{ab}
\phi_b(\tau)-\phi_a(\tau)
i\partial_n X^a(\tau)\right]\right)= \nonumber \\
=\exp\left(-\frac{1}{2}\int d\tau d\tau'
\partial_nX^a(\tau)\triangle(\tau,
\tau')_{ab}^{-1}\partial_n X^b(\tau') \right) \ ,
\triangle (\tau',\tau)^{ab}=i\partial 
\delta (\tau'-\tau)\theta^{ab} \  \nonumber \\
\end{eqnarray}
with
\begin{equation}
\triangle (\tau,\tau')^{-1}_{ab}=
\theta_{ab}\frac{1}{2\pi}
\sum_n\frac{1}{n}e^{in(\tau-\tau')} \ .
\end{equation}
We expand the string field as 
\begin{equation}
X^a(\tau,\rho)=\sqrt{
\frac{\alpha'}{2}}\sum_{n=-\infty,
n\neq 0}^{\infty}\rho^n X^a_ne^{in\tau}, \
\partial_n X^a(\tau,\rho=1)=
\sqrt{\frac{\alpha'}{2}}\sum_{n=-\infty, n\neq 0}^{\infty}
n X^a_n e^{in\tau} \ .
\end{equation}
Note that there is no zero mode thanks to
the Dirichlet boundary conditions.
Then (\ref{Jef}) is equal to
\begin{eqnarray}
-\frac{1}{2}\int d\tau d\tau'
\partial_nX^a(\tau) \triangle(
\tau,\tau')_{ab}^{-1}\partial_n X^b(\tau')=
\alpha'\pi\sum_{m=1}^{\infty}
m \theta_{ab}X^a_{-m}X^b_{m}\ , \nonumber \\
\end{eqnarray}
that  agrees precisely with the expression
\cite{KrausBST}
\begin{equation}
S=\frac{i}{2}\int_0^{2\pi} d\tau F_{ab}X^a(\tau)\dot{X}^b(\tau)=
\pi\alpha'\sum_{n=1}^{\infty}nF_{ab}
X^a_{-n}X^b_n \ , F_{ab}=\theta_{ab} \ ,
\end{equation}
 arising from the term
\begin{equation}
S_A=-i\int_0^{2\pi} d\tau A_a(X^a)\partial_{\tau}X^a(\tau) \ .
\end{equation}
As a result, we have obtained the partition sum on the disk
for a D2p-brane in the presence of the
gauge field $F_{ab}$
\begin{equation}
e^{-I}=\exp \left(i\int d\tau\left[
A_a(X)\partial_{\tau}X^a(\tau)+\frac{\alpha'}{2}
F_{ab}\psi^a(\tau)\psi^b(\tau)\right]\right), 
F_{ab}=\partial_aA_b-\partial_bA_a=\theta_{ab} \ .
\end{equation}
Consequently the action is  equal to
\cite{TseytlinSM,TseytlinDBI} 
\footnote{We consider Minkowski space-time with the
metric $\eta_{\mu\nu}=\mathrm{diag}(-1,1,\dots,1)$.}
\begin{equation}
S(F)=Z(F)=<e^{-I}>=-\sqrt{2}T_{2p}
\int dt d^{2p}x\sqrt{\det (\delta_{ab}+2\pi\alpha'F_{ab})} \ .
\end{equation}
In order to describe  odd dimensional
D-branes we should include the tachyon
in the boundary action and calculate its condensation.
Let us consider the ansatz
\begin{equation}\label{T0}
T=u \Phi_2, \ 
[\Phi_x,\Phi_y]=i\epsilon_{xy}\theta, \ , x,y=1,2, \
 [\Phi_i,\Phi_j]=i\theta_{ij}, \
i, j= 3,\dots, 2p, 
\end{equation}
so that $\theta$ in $[\Phi_a,\Phi_b]=i\theta_{ab}$
has a  form
\begin{equation}
\theta=\left(\begin{array}{ccc}
0 & \theta & 0 \\
-\theta & 0 & 0 \\
0 & 0 & \theta_{ij} \\ \end{array}\right)
\ , i, j=3,\dots, 2p \ .
\end{equation}
Then (\ref{nBPS0})  is equal to 
\begin{eqnarray}
S=<e^{-I}>=
<\int [d\phi_a]\exp
\left(\int d\tau\left[ i\frac{1}{2}
\phi_a\theta^{ab}\dot{\phi}_b
+\right.\right. \nonumber \\
\left.\left.
-\alpha'u^2\theta^2\psi^1\frac{1}{\partial_{\tau}}
\psi^1
-u^2(\phi_2)^2
+i\phi_a\partial_nX^a
+\frac{i\alpha'}{2}\theta_{ab}
\psi^a\psi^b\right]\right)> \ . \nonumber \\
\end{eqnarray}
 We can take out the following expression
from the path integral over $\phi$
\begin{equation}
\exp\left(\int d\tau\left[\alpha'
u^2\theta^2\psi^1\frac{1}{\partial_{\tau}}
\psi^1 +\frac{i\alpha'}{2}\theta_{ab}
\psi^a\psi^b\right]\right) \ .
\end{equation}
Next calculation is the same as 
the case with vanishing tachyon field.
The  $i,j=3,\dots, 2p$ components  
give the
same result as in the case of even dimensional
D-brane without exciting tachyon and
the integration over $x,y=1,2$ gives
\begin{eqnarray}\label{up}
\exp\left(-\frac{1}{2}\int d\tau d\tau'
\partial_nX^x(\tau)
\triangle(\tau-\tau')_{xy}^{-1}\partial_nX^y
(\tau')\right) \ , \nonumber \\
\triangle(\tau-\tau')_{xy}^{-1}=
\sum_n (E_n)_{xy}^{-1}e^{in(\tau-\tau')}, \ 
(E^{-1}_n)_{xy}=\frac{\theta^2}{n^2}\left(
\begin{array}{cc} -\frac{1}{2}u^2 & -\frac{n}{\theta} \\
\frac{n}{\theta} & 0 \\ \end{array}\right) \ , \nonumber \\
\end{eqnarray}
which together with the $F_{ij}$ and  $F_{xy}$ terms
gives the result
\cite{MooreBT2,Aleksejev}
\begin{eqnarray}
S=<e^{-I}>=
<\exp \left(
\int d\tau\left[-(u^2 (X^1)^2+
\alpha'u^2\psi^1\frac{1}{\partial_{\tau}}
\psi^1)\right.\right. + \nonumber \\
\left.\left.+iA_a(X^a)\partial_{\tau}X^a+
\frac{i\alpha'}{2}F_{ab}\psi^a\psi^b\right]
\right)> \ ,F_{ab}=\theta_{ab} \ , \nonumber \\
\end{eqnarray}
where we have made replacement $u\theta\rightarrow u$.
Then it is easy to see that the tachyon condensation
really leads to the emergence of odd dimensional
D-brane exactly in the same way as in \cite{MooreBT2}.
The partition function is equal to \cite{MooreBT2,
Aleksejev,Yasnov}
\begin{equation}
Z=K\int dt d^{2p-1}x
Z(a,v)_{fermi}
\sqrt{\det (\delta_{ij}
+2\pi\alpha'F_{ij})} \ ,
v=\frac{4\pi\alpha' u^2}{1+(2\pi\alpha' \theta)^2}  \ .
\end{equation}
with \cite{MooreBT2} 
\begin{equation}
Z(a,v)_{fermi}=4^v \frac{Z_1(v)^2}{Z_1(2v)} \ ,
Z_1(v)=\sqrt{v}e^{\gamma v}\Gamma(v) \ ,
\end{equation}
 and where $K$ is a numerical factor that will be
determined as in \cite{KlusonBS}.
When we calculate the partition sum for constant
tachyon $T=a1_{N\times N}$ we obtain the exact
tachyon potential $e^{-2\pi a^2}$ multiplied 
with the DBI term $\sqrt{\det(\delta_{ab}
+2\pi\alpha'F_{ab})}$ arising from the partition sum
calculated in the pure gauge field background.
 Then we can expect that for a slowly
varying tachyon field the action corresponds to the
non-BPS D2p-brane action 
\begin{equation}\label{Spom}
S=-\sqrt{2}T_{2p}\int dt d^{2p}x e^{-2\pi T^2}\sqrt{
(1+(2\pi\alpha' \theta)^2)\det (\delta_{ij}+
2\pi\alpha' F_{ij})}+O(\partial T) \ ,
\end{equation}
This action  evaluated on the tachyon profile $T=ux^1$ should
be  equal to the partition sum in the limit $u\rightarrow 0$.
 From this requirement we
can determine the overall  normalisation constant $K$.
 Then 
the  action (\ref{Spom}) evaluated on the
tachyon profile  $T(x)=ux^1$ is equal to 
\begin{eqnarray}
-\sqrt{2}T_{2p}\int dtd^{2p-1}x
\sqrt{(1+(2\pi\alpha' \theta)^2)\det (\delta_{ij}+
2\pi\alpha' F_{ij})}\int dx^1 e^{-2\pi u^2(x^1)^2}=\nonumber \\
=-\frac{1}{u}T_{2p}
\int d^{2p-1}x\sqrt{(1+(2\pi\alpha' \theta)^2)\det (\delta_{ij}+
2\pi\alpha' F_{ij})} \ . \nonumber \\
\end{eqnarray}
On the other hand, in the limit $u\rightarrow 0$ we have
\begin{equation}
Z(v)_{fermi}=\
 \sqrt{\frac{2}{v}}+O(v) , \ v\sim 0 \ ,
\end{equation}
so that the partition function is equal to
\begin{equation}
S=K\frac{1}{\sqrt{2\pi \alpha'u^2}}\int dt d^{2p-1}x
\sqrt{(1+(2\pi\alpha' \theta)^2)\det (\delta_{ij}+
2\pi\alpha' F_{ij})}   
\end{equation}
and consequently
\begin{equation}
K=-T_{2p}\sqrt{2\pi\alpha'} \ .
\end{equation}
Using this result it is easy to see that the action
arising from the tachyon condensation is
equal to (In this case the tachyon condensation
 corresponds to $u\rightarrow \infty$
\cite{MooreBT2})
\begin{eqnarray}\label{Sfinal}
S=Z(\infty)=-
\sqrt{2\pi\alpha'}T_{2p}\sqrt{2\pi}\int
dt d^{2p-1}x\sqrt{\det (\delta_{ij}+
2\pi\alpha' F_{ij})}= \nonumber \\
=-T_{2p-1}\int dt d^{2p-1}x
\sqrt{\det(\delta_{ij}+2\pi\alpha'F_{ij})}\ ,
\nonumber \\
\end{eqnarray}
where we have used
\begin{equation}\label{nBPSZ}
Z(v)\sim \sqrt{2\pi}+O(v^{-1}) \ , u\rightarrow \infty \ .
\end{equation} 
The result (\ref{Sfinal}) is a correct value of the
action for D(2p-1)-brane with the 
background gauge field strength $F_{ij}$. However,
we must stress one important thing. It seems
that in this approach we cannot construct 
space-time filling D9-brane as can be seen
from (\ref{T0}). On the other hand  we can  construct
the space-time filling unstable \dd9 system with
equal gauge field background.  Let us
consider formally the same ansatz as (\ref{T0})
 \begin{equation}\label{T9}
T=u \Phi_{10}, \ 
[\Phi_x,\Phi_y]=i\epsilon_{xy}\theta, \ , x,y=9, 10, \
 [\Phi_i,\Phi_j]=i\theta_{ij}, \
i, j= 1,\dots, 8, 
\end{equation}
so that $\theta$ in $[\Phi_a,\Phi_b]=i\theta_{ab}$
has a  form
\begin{equation}\label{theta9}
\theta=\left(\begin{array}{ccc}
\theta_{ij} & 0 & 0 \\
0  & 0 & \theta \\
0 & -\theta & 0  \\ \end{array}\right)
\ , i, j=1,\dots, 8 \ .
\end{equation}
Now $\Phi_{10}$ is quantum mechanical operator
without any space-time meaning. We see that we 
can perform the same calculation as above with the
result
\begin{eqnarray}\label{nBPS9}
S=<e^{-I}>=
<\exp \left(
\int d\tau\left[-(u^2(X^9)^2+
\alpha'u^2\psi^9\frac{1}{\partial_{\tau}}
\psi^9)\right.\right. + \nonumber \\
\left.\left.+iA_i(X^i)\partial_{\tau}X^i+
\frac{i\alpha'}{2}F_{ij}\psi^i\psi^j\right]
\right)> \ ,F_{ij}=\theta_{ij} \ , \nonumber \\
\end{eqnarray}
where we have used the fact that there is no
term proportional to $\partial_n X^{10}$. Naively
we could say that this action corresponds to
space-time filling D9-brane. However we know that
 D9-brane is stable so that it does not contain
 tachyon in its spectrum. Moreover, we know that 
D8-brane that arises from the tachyon condensation
in the previous expression in the limit
$u\rightarrow \infty$ is unstable so that its tension is $\sqrt{2}T_8$.
We also know that in the limit $u\rightarrow \infty$
we get the numerical value (\ref{nBPSZ}). From these
facts we conclude that the normalisation constant
in (\ref{nBPS9}) is equal to $2T_{9}$. In other words,
the action (\ref{nBPS9}) corresponds to the partition
sum for \dd 9 system with equal gauge field 
background on their world-volumes as it will be 
clear from the analysis in the next section. 

In order to describe single D9-brane in this approach we
propose other process which unfortunately does not
seem to have simple description in BSFT. We 
construct an infinite number of D7-branes through the
tachyon condensation and then single D9-brane arises
by ordinary matrix theory construction reviewed above.
 Unfortunately,
the tachyon condensation in BSFT describing general
configuration of D-branes seems to be difficult problem
since the boundary theory is not free \cite{MooreBT}.
For that reason we leave this issue for the future research.

\section{\dd-system}
\label{fourth}
 The action proposed in (\ref{second})
section corresponds to the space-time 
 filling  \dd 9 system, for lower dimensional
ones with $x^{\mu}, \mu=0,\dots, p$ Neumann boundary
conditions and $x^i, \ i=p+1,\dots, 9$ Dirichlet boundary conditions
the matrix $M$ has a form
\begin{equation}
M(A,\Phi,A',\Phi',T)=
\left(\begin{array}{cc}-
iA_{\mu}({\bf X^{\mu}})D{\bf X}^{\mu}-
i\Phi_i({\bf X}^{\mu})\tilde{D}{\bf X}^i &
 \overline{T}({\bf X}^{\mu}) \\
T({\bf X}^{\mu}) & 
-iA_{\mu}({\bf X}^{\mu})D{\bf X}^{\mu}
-i\Phi_i'({\bf X}^{\mu})\tilde{D}
{\bf X}^{i}\\ \end{array}\right) \ ,
\end{equation}
where now $\Phi_i,\Phi'_i$ describe transverse
fluctuations of $N$ Dp-branes and $
\overline{D}$p-branes
respectively and where $\tilde{D}=\partial_{\theta}
+\theta\partial_n $.

In the following we will consider 
 $N$ \dd 0 system  in the limit $N\rightarrow \infty$.
 Now the gauge fields
$A_0,A_0'$ are not dynamical fields and we
can eliminate them thanks to the gauge invariance.
Then $M$ has a form
\begin{equation}
M(\Phi,\Phi',T)=
\left(\begin{array}{cc}
-i\Phi_I({\bf X}^{0})\tilde{D}{\bf X}^I &
 \overline{T}({\bf X}^{0}) \\
T({\bf X}^{0}) & 
-i\Phi_I'({\bf X}^{0})\tilde{D}
{\bf X}^{I}\\ \end{array}\right) ,\ 
I=1,\dots, 9 \ .
\end{equation}
Performing the same calculation as in
(\ref{second}) section we obtain T-dual version of
the action (\ref{I13}) (We consider time-independent
background fields) 
\begin{eqnarray}\label{D0d}
S(\Phi,\Phi',T)=<\exp\left(-I(\Phi,\Phi',T)\right)> \ ,  
I=\int d\tau \left(\overline{\eta}_a
\partial_{\tau}\eta^a+\overline{\eta}_a
\mathcal{M}^a_b\eta^b\right) \ ,
 \nonumber \\
\mathcal{M}=\left(\begin{array}{cc}
-i(\Phi_I\partial_nX^I-i\frac{\alpha'}{2}
[\Phi_I,\Phi_J]\psi^I\psi^J) +\overline{T}T &
i\sqrt{\alpha'}(\Phi_I\overline{T}-\overline{T}\Phi'_I)\psi^I \\
-i\sqrt{\alpha'}(T\Phi_I-\Phi_I'T)\psi^I &
-i(\Phi_I\partial_nX^I-i\frac{\alpha'}{2}
[\Phi_I,\Phi_J]\psi^I\psi^J )+
T\overline{T} \\ \end{array}\right)
 \ . \nonumber \\
\end{eqnarray}
To begin with, consider the configuration with the tachyon
fields equal to the constant value $T$ and with all D0-branes
sitting in the origin which corresponds to $\Phi^I,\Phi'^I=0$.
Then 
\begin{equation}
\mathcal{M}=\left(\begin{array}{cc}
\overline{T}T & 0 \\
0 & T\overline{T} \\ \end{array}\right) \ .
\end{equation}
Note that now the tachyon field is on the main diagonal
so that it should be regarded as a field with even grading.
 Then we obtain
\cite{KrausBST}
\begin{equation}
S=2T_0\tr_{N\times N} e^{-2\pi T\overline{T}} \ .
\end{equation}
We see
that the unstable \dd 0-system corresponds to
the tachyon value $T=0$ and the closed string vacuum
to the value $T=\infty$. 
It is also clear 
 that we can construct many  configurations
around the perturbative unstable vacuum $T=0$,
  for example
\begin{equation}
[\Phi^1_i,\Phi^1_j]=i\theta_{ij}^1, \ i,j=1,\dots, 2p_1, \
[\Phi^2_i,\Phi^2_j]=i\theta_{ij}^2, \ i,j=1,\dots, 2p_2 
\end{equation}
with the first commutator corresponding to D0-branes and the
second one to $\overline{D}$0-branes.
When we insert this ansatz into (\ref{D0d}) we see  that
$\mathcal{M}$ is again block diagonal and consequently
the partition function reduces to two independent parts corresponding
to the partition sum for $D2p_1$-brane and $D2p_2$-brane
respectively
\begin{equation}
S_{D2p_1+D2p_2}=
\sum_{k=1}^2<\tr P\exp\left(i\int d\tau 
\left[\Phi_i^k\partial_nX^i+\frac{\alpha'}{2}
\theta_{ij}^k\psi^i\psi^j\right]\right)> \ .
\end{equation}
Each  partition function is equivalent to the partition
function for $D2p_{1,2}$-brane with the background gauge field
$\theta^{1,2}_{ij}$ as we have shown in \cite{KlusonBS}.
As we can expect from the general theory \cite{WittenK}
this configuration
either annihilates to the closed string vacuum or to the state
with the lower dimensional D-brane charge. The final state
of the configuration  depends on the
form of the background field strengths $F^{1,2}=\theta^{1,2}$. 
It would be nice to study this process in the BSFT theory 
however it seems to be difficult task since the world-sheet theory
is not free even in the case of the constant tachyon field.
On the other hand, in \cite{KrausBST,TakayBT} the lower dimensional
D-branes emerge naturally from \dd \ system even in the  
case of the vanishing gauge field. We can then expect that
higher dimensional D-branes arise from \dd 0 \ system in
the similar manner.
 
To see directly the emergence of D-branes we should turn on
the tachyon field. We restrict ourselves on the linear
profile of the tachyon  as in \cite{KrausBST}. Let us consider
the ansatz
\begin{equation}\label{ans1}
[\Phi_a,\Phi_b]=i\theta_{ab},  \ a, \ b =1,\dots,2p,
\ [\Phi'_i,\Phi_j']=i\theta_{ab} , \ a,b=1,\dots, 2p, \
T=u^a\Phi_a , 
\end{equation}
where $u^a$ are complex numbers. In other words, we presume
that $\Phi,\Phi'$ are the same quantum mechanical operators.  
Then $\mathcal{M}$ is equal to
\begin{equation}
\mathcal{M}=
\left(\begin{array}{cc}
-i(\Phi_a\partial_nX^a+\frac{\alpha'}{2}
\theta_{ab}\psi^a\psi^b)+
u^{*a}u^b\Phi_a\Phi_b & \sqrt{\alpha'}\theta_{ab}u^{a*}\psi^b \\
\sqrt{\alpha'}\theta_{ab}u^a\psi^b & 
-i(\Phi_a\partial_nX^a+
\frac{\alpha'}{2}\theta_{ab}\psi^a\psi^b)+
u^{a*}u^b\Phi_a\Phi_b \\ 
\end{array}\right) \ .
\end{equation}
 We  rewrite the path integral
over $\eta$ as the path-ordered trace. 
Then as in (\ref{second}) section we can
write $\mathcal{M}$ as
\begin{equation}
\mathcal{M}=1_{2\times 2}
\otimes \mathcal{M}_1(\Phi)_{N\times N}+
\mathcal{M}_{2,2\times 2}\otimes 1_{N\times N} \ ,
\end{equation}
Now we can easily rewrite the first term in the form
of the path integral over $\phi_a$ so we obtain
\begin{eqnarray}
<e^{-I}>=
<\tr_{2\times 2} P\int [d\phi_a]
\exp\left(\int d\tau\left[1_{2\times 2}
\otimes ( i\frac{1}{2}\phi_a\theta^{ab}\dot{\phi}_b
-u^{a*}u^b\phi_a\phi_b+i\phi_a\partial_nX^a)
\right.\right. - \nonumber \\
\left.\left.
-\sqrt{\alpha'}\theta_{ab}
\left(\begin{array}{cc} 
0 & u^{a*}\psi^b \\
u^a\psi^b & 0 \\ \end{array}\right)
+1_{2\times 2}\otimes 
i\frac{\alpha'}{2}\theta_{ab}
\psi^a\psi^b\right]\right)> \ .\nonumber \\
\end{eqnarray}
We propose  the ansatz
\begin{equation}\label{DDans}
T=\sum_{x=1}^4u^xX_x , \ u^1=u_1, u^2=u^4=0,
 \ u^3=iu_3, \,
[\Phi_a,\Phi_b]=i\theta_{ab}, \ a, \ b=
1,\dots,2p , \ 
\end{equation}
with 
\begin{equation}
\theta_{ab}=\left(\begin{array}{ccccc}
0 & \theta_1 & 0 & 0 & 0 \\
-\theta_1 & 0 & 0 & 0 & 0 \\
0 & 0 & 0 & \theta_2 & 0 \\
0 & 0 & -\theta_2 & 0 & 0 \\
0 & 0 & 0 & 0 & \theta_{ij} \\ \end{array}\right) ,
\ i, j=5,\dots, 2p  \ .
\end{equation}
We must stress that this is not
the most general ansatz, rather this is the background
for which we can explicitly calculate the
BSFT action. 
Then 
\begin{equation}
u_iu_j^* \phi^i\phi^j 
\rightarrow 
u_1^2\phi^2_1+u_3^2\phi_3^2 
\end{equation}
and also
\begin{equation}
(E_n)_{xy}=\left(
\begin{array}{cccc}
-u_1^2 & \frac{n}{\theta_1} & 0 & 0 \\
-\frac{n}{\theta_1} & 0 & 0 & 0 \\
0 & 0 & -u_2^2 & \frac{n}{\theta_2} \\
0 & 0 & -\frac{n}{\theta_2} & 0 \\ \end{array}\right), \ 
\det(E_{xy})=\frac{n^2}{\theta_1^2}\frac{n^2}{\theta_2^2}, \ 
(E_n)^{-1}_{xy}=\left(\begin{array}{cccc}
0 & -\frac{\theta_1}{n} & 0 & 0 \\
\frac{\theta_1}{n} & -\frac{\theta_1^2u_2^2}{n^2} 
& 0 & 0 \\
0 & 0 & 0 & -\frac{\theta_2}{n} \\
0 & 0 & \frac{\theta_2}{n} & 
-\frac{\theta_2^2u^2_2}{n^2} \\ \end{array}\right) \ .
\end{equation}
Terms proportional to $\theta_{ij}$ give 
the standard result reviewed in previous parts.
On the other hand, for $x,y=1,\dots,4$ we obtain 
\begin{eqnarray}
-\frac{1}{2}\int d\tau d\tau'\partial_n X^x \triangle 
(\tau-\tau')_{xy}^{-1}\partial_n X^y(\tau')=
\frac{\pi\alpha'}{2}\sum_n\theta_{xy}nX_{-n}^xX^y_n-\nonumber \\
-\frac{\pi\alpha'}{2}\sum_{n, n\neq 0}
\left(\tilde{u}_2^2X_{-n}^2X_n^2+\tilde{u}_4^2
X_{-n}^4X_n^4\right)
=\frac{i}{2}\int_0^{2\pi}d\tau \theta_{xy}
X^x(\tau)\dot{X}^y(\tau)-\int_0^{2\pi}
d\tau T(X)^2 \ , \nonumber 
\\ T(X)=\tilde{u}_2^2(X^2(\tau))^2+
\tilde{u}_4^2(X^4(\tau))^2 , \
\tilde{u}_2=\theta_{1}u_1, \
\tilde{u}_4=\theta_2 u_3 \ ,\nonumber \\
\end{eqnarray}
so we have
\begin{eqnarray}
<e^{-I}>=
<\tr_{2\times 2} P\exp\left(\int d\tau
\left[1_{2\times 2}\otimes
(iA_a(X)\partial_{\tau}X^a+\frac{i\alpha'
F_{ab}}{2}
\psi^a\psi^b-T(X)^2)-\right.\right.\nonumber \\
\left.\left.-\sqrt{\alpha'}
\left(\begin{array}{cc} 
0 & \theta_{12}u^1\psi^2-i
\theta_{34}u^3\psi^4 \\
\theta_{12}u^1\psi^2+i
\theta_{34}u^3\psi^4 & 0  \\ \end{array}
\right) \right]\right)>=<\exp\left(\int d\tau\left[
iA_a(X)\partial_{\tau}X^a+\right.\right. \nonumber \\
\left.\left. +\frac{i\alpha'F_{ab}}{2}
\psi^a\psi^b-T(X)^2-\overline{\xi}
\dot{\xi} 
-(\xi+\overline{\xi}) \sqrt{\alpha'}\tilde{u}_2\psi^2
-i\sqrt{\alpha'}(\overline{\xi}-\xi)
\tilde{u}_4\psi^4)\right]\right)>\ , \nonumber \\
\end{eqnarray}
where we have used the correspondence between
fermionic fields $\xi,\overline{\xi}$ and Pauli matrices
which was explicitly demonstrated
in  (\ref{second}) section.
It is now  easy to study the tachyon condensation,
following \cite{KrausBST,TakayBT}. Firstly we integrate 
out the fermionic field $\xi, \overline{\xi}$ which gives 
\begin{equation}
\xi=-\sqrt{\alpha'}\frac{1}{\partial_{\tau}}
(u_2\psi^2+iu_4\psi^4) \ ,
\overline{\xi}=-\sqrt{\alpha'}
\frac{1}{\partial_{\tau}}(u_2\psi^2-iu_4\psi^4) \ ,
\end{equation}
where we have renamed $\tilde{u}_x\rightarrow u_x$.
Consequently we have
\begin{eqnarray}\label{Sbon}
<e^{-I}>=
<\exp\left(\int d\tau\left[
iA_a\partial_{\tau}X^a+\frac{i\alpha'}{2}
F_{ab}\psi^a\psi^b-\right.\right. \nonumber \\
\left.\left. -T(X)^2
-\alpha'
\left(u_2^2\psi^2
\frac{1}{\partial_{\tau}}\psi^2+u^2_4\psi^4
\frac{1}{\partial_{\tau}}\psi^4\right)\right]\right)> \ . \nonumber \\
\end{eqnarray}
Now the calculation is the same as in
the original papers \cite{MooreBT2,KrausBST,
TakayBT}.   Thanks to the
special form of the ansatz 
(\ref{ans1},\ref{DDans}) the partition sum
factorises into the subspaces labelled with the coordinates
$x^{1,2}$, $x^{3,4}$ and $x^{5,\dots,2p}$.
Then we can use the approach outlined in the
previous section with the result
\begin{eqnarray}\label{Sbon1}
S=K\int dt d^{2p-2}x Z(v_1,v_2)e^{-2\pi
T\overline{T}}\sqrt{\det (\delta_{ij}+
2\pi\alpha' F_{ij})} \ , \nonumber \\
Z(v_1,v_2)= \prod_{i=1}^2
4^{v_{2i}}\frac{Z_1(v_{2i})^2}{Z_1(2v_{2i})} \ ,
Z_1(x)=\sqrt{x}e^{\gamma x}\Gamma (x) \ ,
v_{2i}=\frac{4\pi\alpha' u_{2i}^2}{1+(2\pi\alpha' \theta_i)^2} \ . \nonumber \\
\end{eqnarray} 
As usual we can expect that for a slowly varying tachyonic field 
this action reduces to the sum of two DBI actions for a D-brane and
an anti-D-brane respectively. However thanks to the particular
choose of the background fields these two actions are equal so
that we  expect that the action will have a form
\begin{eqnarray}
S=-2T_{2p}\int dt d^{2p}x e^{-2\pi T\overline{T}}
\sqrt{\det (\delta_{ab}+2\pi\alpha' F_{ab})}+O(\partial T)=\nonumber \\
=-2T_{2p}\int dt d^{2p}xe^{-2\pi T\overline{T}}
\sqrt{(1+(2\pi\alpha' \theta_1)^2)
(1+(2\pi\alpha' \theta_2)^2)}\sqrt{\det (\delta_{ij}+
2\pi\alpha' F_{ij})}+O(\partial T)  \ . \nonumber \\
\end{eqnarray}
As in the previous section we should work out this action on
the tachyon profile $ T\overline{T}= u_2^2(x^2)^2+
u_4^2(x^4)^2$ which gives  
\begin{eqnarray}
S=-2T_{2p}\int dt d^{2(p-1)}x
\sqrt{(1+(2\pi\alpha' \theta_1)^2)
(1+(2\pi\alpha' \theta_2)^2)}\sqrt{\det (\delta_{ij}+
2\pi\alpha' F_{ij})}\times \nonumber \\
\times \int dx^2e^{-2\pi u_2^2 (x^2)^2}
\int dx^4 e^{-2\pi u_4^2 (x^4)^2}=-
\frac{T_{2p}}{u_2u_4}\int dt d^{2p-2}x
\sqrt{\det (\delta_{ab}+2\pi\alpha' F_{ab})} \ .
\nonumber \\
\end{eqnarray}
On the other hand, for $u_i \sim 0$ we 
have $Z(v_1,v_2)\sim \frac{2}{
\sqrt{v_2v_4}} $
and consequently (\ref{Sbon1}) is equal to 
\begin{equation}
S=\frac{K}{2\pi\alpha' u_2u_4}\int 
dt d^{2p-2}x\sqrt{((1+(2\pi\alpha' \theta_1)^2)
(1+(2\pi\alpha' \theta_2)^2)}
\sqrt{\det (\delta_{ij}+2\pi\alpha' F_{ij})} 
\end{equation}
Comparing these two expressions we obtain 
the value of the normalisation constant
\begin{equation}
K=-2\pi\alpha' T_{2p}.
\end{equation}
Then it is easy to study the tachyon condensation.
For $u_2\rightarrow \infty $ and $u_4\rightarrow \infty$
we have
\begin{equation}
4^v\frac{Z_1(v)^2}{Z_1(2v)}\sim
\sqrt{2\pi} , v\rightarrow \infty
\end{equation}
and consequently the action is equal to
\begin{equation}
-T_{2p-2}\int dtd^{2p-2}x\sqrt{\det(\delta_{ij}
+2\pi\alpha' F_{ij})}
\end{equation}
which is the correct value of the action
 for $D_{2p-2}$-brane with the background
gauge field $F_{ij}$. 
On the other hand, for $u_2\rightarrow 0, u_4\rightarrow \infty$ we
have
\begin{equation}
Z(v_1,v_2)\rightarrow 
\frac{\sqrt{4\pi}}{\sqrt{v_2}}=\frac{\sqrt{1+(2\pi\theta_1)^2}}{
\sqrt{\alpha'}u_2}
\end{equation}
so that
\begin{eqnarray}
S=-\frac{\sqrt{2}T_{2p-1}}{\sqrt{2}u_2}
\int dt d^{2p-2}x \sqrt{(1+(2\pi\alpha'\theta_1)^2)
\det(\delta_{ij}+2\pi\alpha' F_{ij})}=\nonumber \\
=-\sqrt{2}T_{2p-1}\int dt dx^1 e^{-2\pi u_2^2(x^2)^2}
d^{2p-2}x\sqrt{(1+(2\pi\alpha' \theta_1)^2)
\det (\delta_{ij}+2\pi\alpha'F_{ij})}= \nonumber \\
=-\sqrt{2}T_{2p-1}\int dt d^{2p-1}x\sqrt{(1+
(2\pi\alpha'  \theta_1)^2)\det(\delta_{ij}+
2\pi\alpha'F_{ij})} \nonumber \\
\end{eqnarray}
which is a correct action for non-BPS D(2p-1)-brane with
the background gauge field $F_{12}=\theta_1, F_{ij}$.
To conclude, we have shown in this section that Dp-branes,
$p\leq 6$ 
naturally arise  from $N$ D0 and $\overline{D}0$-branes
in the limit $N\rightarrow \infty$ through
the particular ansatz (\ref{DDans}). We also
see that (\ref{DDans}) cannot lead to
the emergence of D7, D8 and D9-brane since
the maximal number of $2p=8$. 
In order to describe these D-branes we should proceed
in the same way as was outlined in the end of
the previous section. In order  to describe non-BPS D9-brane
we take the ansatz
\begin{eqnarray}\label{ans2}
[\Phi_a,\Phi_b]=i\theta_{ab},  \ a, \ b =1,\dots,8,
\ [\Phi'_i,\Phi_j']=i\theta_{ab} , \ a,b=1,\dots,8, \nonumber \\
T=u\Phi_{10}, \ u=u^*, 
\ [\Phi_9,\Phi_{10}]=i\theta  , \nonumber \\
\end{eqnarray}
with $\theta$ the same as in (\ref{theta9}).
When we insert this configuration into (\ref{D0d})
$\mathcal{M}$ is equal to
\begin{equation}
\mathcal{M}=
\left(\begin{array}{cc}
-i(\Phi_a\partial_nX^a+\frac{\alpha'}{2}
\theta_{ab}\psi^a\psi^b)+
u^2\Phi_{10}^2 & \sqrt{\alpha'}\theta u\psi^9 \\
\sqrt{\alpha'}\theta u\psi^9 & 
-i(\Phi_a\partial_nX^a+
\frac{\alpha'}{2}\theta_{ab}\psi^a\psi^b)+
u^2\Phi_{10}^2 \\ 
\end{array}\right) \ .
\end{equation}
We see that the previous expression is a special form
of $\mathcal{M}$  which we have discussed in
(\ref{second}) section and which leads to the partition
sum for non-BPS D-brane through the $(-1)^{F_L}$
orbifold projection.
Then the previous expression naturally leads
to the partition function for  D9-brane
whose boundary theory has two fix points: one corresponding
to non-BPS D9-brane and the second one to
BPS D8-brane. We see that D8 and D9-branes
 naturally arise from
the \dd 0 system. And finally,  to describe D7-brane
we can use a modification of (\ref{D0d}) where we 
replace one part of the tachyon field (say $T=u_3\Phi^3$)
with $T=u_9 \Phi^{10}$ with the commutation relations
between various $\Phi$' the same as in (\ref{ans2}). Then
it is easy to see that the limit $u_1,u_9\rightarrow \infty$
leads to D7-brane exactly in the same way as we have shown
in previous parts. 
 We can then claim
that all D-branes naturally emerge from \dd 0 system.

\section{Conclusion}\label{fifth}
In this paper we have studied the 
emergence of D-branes from $N$ of \dd 0
system  in type IIA theory, following
our recent paper \cite{KlusonBS}. 
We have seen that the \dd-system is very interesting
and seems to be the most general unstable
configuration in  type IIA, IIB theories. Many
problems and issues with these configurations have
been considered recently in the papers
\cite{KrausMT,MandalMT,LiMT}. In
this paper we have tried to show that
the BSFT theory is also very useful tool
for study the emergence of  higher dimensional
D-branes from lower dimensional ones.

However, many open question remains. We have
seen that there is a problem with the tachyon
condensation on the \dd-system with
different gauge fields background. From the
general arguments regarding tachyon condensation
\cite{SenR,WittenK} we can expect that after the
tachyon condensation either lower
dimensional D-brane emerges or the system ends
in the closed string vacuum according
to the value of background gauge field. It
would be nice to study this process directly in
BSFT theory. It would be also nice to study
the fluctuations around various nontrivial 
configurations which arise from the tachyon
condensation in the BSFT approach. And
finally, it should be addressed the question
of the relation BSFT to the Matrix theory.
\\
\\
{\bf Acknowledgements}
We would like to thank Rikard von Unge for very helpful 
discussions. This work was supported by the
Czech Ministry of Education under Contract No.
143100006.

\end{document}